\documentclass[aps,prx,twocolumn,superscriptaddress]{revtex4-2}
\usepackage[utf8]{inputenc}
\usepackage{longtable}
\usepackage{amsmath}%
\usepackage{amsfonts}%
\usepackage{amssymb}%
\usepackage{graphicx}
\usepackage[hidelinks, breaklinks=true]{hyperref}
\usepackage{CJKutf8} 
\usepackage{ulem} 
\usepackage[product-units=power, separate-uncertainty=true, multi-part-units=single]{siunitx} 
\usepackage{xcolor}

\newcommand{\fig}{Fig.~\ref}

\newcommand{\SrBiSe}{Sr$_{x}$Bi$_2$Se$_3$}
\newcommand{\CuBiSe}{Cu$_x$Bi$_2$Se$_3$}
\newcommand{\BiSe}{${\mathrm{Bi}}_{2}{\mathrm{Se}}_{3}$}
\newcommand{\Fc}{F_\mathrm{c}^{H}}
\newcommand{\Pc}{P_\mathrm{c}^{H}}

\newcommand{\myref}[7]{\href{http://dx.doi.org/#7}{#1, #2, #3 \textbf{#4}, #5 (#6).}} 
\newcommand{\textref}[1]{{#1}} 

\newcommand{\etal}{\textit{et al.}}

\newcommand{\pgi}{Peter Grünberg Institut (PGI-3), Forschungszentrum Jülich, 52425 Jülich, Germany}
\newcommand{\jara}{Jülich Aachen Research Alliance (JARA), Fundamentals of Future Information Technology, 52425 Jülich, Germany}
\newcommand{\rwth}{Experimentalphysik IV A, RWTH Aachen University, Otto-Blumenthal-Straße, 52074 Aachen, Germany}
\newcommand{\koeln}{Physics Institute II, University of Cologne, 50937 Köln, Germany}
\newcommand{\diam}{Diamond Light Source Ltd, Didcot, OX110DE, Oxfordshire, United Kingdom}

\begin{document}

\title{The Vertical Position of Sr Dopants in the Sr$_x$Bi$_2$Se$_3$ Superconductor}
\author{You-Ron~Lin (\begin{CJK*}{UTF8}{bsmi}林又容\end{CJK*})}
                            \affiliation{\pgi} \affiliation{\jara} \affiliation{\rwth}
\author{Mahasweta~Bagchi}   \affiliation{\koeln}
\author{Serguei~Soubatch}   \affiliation{\pgi} \affiliation{\jara}
\author{Tien-Lin~Lee (\begin{CJK*}{UTF8}{bsmi}李天麟\end{CJK*})}  \affiliation{\diam}
\author{Jens~Brede}         \email{brede@ph2.uni-koeln.de (he/him/his) and c.kumpf@fz-juelich.de (he/him/his)}               \affiliation{\koeln}
\author{Fran\c{c}ois~C.~Bocquet}    \affiliation{\pgi} \affiliation{\jara}
\author{Christian~Kumpf}    \email{brede@ph2.uni-koeln.de (he/him/his) and c.kumpf@fz-juelich.de (he/him/his)}
                            \affiliation{\pgi} \affiliation{\jara} \affiliation{\rwth}
\author{Yoichi~Ando }       \affiliation{\koeln}
\author{F.~Stefan~Tautz}    \affiliation{\pgi} \affiliation{\jara} \affiliation{\rwth}

\date{\today}

\begin{abstract}
  The discovery of topological superconductivity in doped \BiSe{} made this class of materials highly important for the field of condensed matter physics. However, the structural origin of the superconducting state remained elusive, despite being investigated intensively in recent years. We use scanning tunneling microscopy and the normal incidence x-ray standing wave (NIXSW) technique in order to determine the vertical position of the dopants -- one of the key parameters for understanding topological superconductivity in this material -- for the case of \SrBiSe. In a novel approach we analyze the NIXSW data in consideration of the inelastic mean free path of the photoemitted electrons, which allows us to distinguish between symmetry equivalent sites. We find that Sr-atoms are not situated inside the van der Waals gap between the \BiSe{} quintuple layers but rather \textit{in} the quintuple layer close to the outer Se planes.
\end{abstract}

\maketitle

\section{Introduction}

Shortly after the discovery of superconductivity in Cu-doped \BiSe{}\ crystals \cite{Hor2010}, Fu and Berg\ \cite{Fu2010} proposed that any electron-doped \BiSe{} is indeed a viable candidate for hosting topological superconductivity with spin-triplet-like pairing.
The spin-triplet-like nature of the pairing was successively confirmed by temperature-dependent nuclear magnetic resonance Knight shift ($K_\mathrm{S}$) experiments\ \cite{Matano2016} which found no change in $K_\mathrm{S}$ below $T_\mathrm{c}$ for magnetic fields applied parallel to the c-direction.
Moreover, the same experiments found that the threefold symmetric \BiSe{} lattice showed a two-fold anisotropy of $K_\mathrm{S}$ when the magnetic field was rotated in the a-b plane.
Such a two-fold symmetry was successively also observed in scanning tunneling microscopy (STM)~\cite{Tao2018}, specific heat\ \cite{Yonezawa2017}, or high-resolution x-ray diffraction (XRD) experiments\ \cite{Kuntsevich2018}.
Theoretically, this breaking of the rotational symmetry in doped \BiSe{} due to an anisotropy of the superconducting gap amplitude is only compatible with nematic superconductivity\ \cite{Yonezawa2019}, which must be topologically non-trivial. 
Since the superconducting state of this material class has been extensively characterized, it is surprising that the structural characterization of the dopant location remains comparatively poor.

Theoretical studies based on density functional theory (DFT) show that doping by Cu~\cite{Wang2011} and Sr~\cite{Li2018} atoms between \BiSe{} quintuple layers is energetically most favorable. However, no direct experimental observations of dopants in the van der Waals (vdW) gap have been reported to date.
An expansion of the c-axis in \SrBiSe{}~\cite{Shruti2015,Liu2015} as well as in \CuBiSe{}~\cite{Shirasawa2014,Wang2016,Froehlich2020} was observed with XRD and was interpreted as an evidence for dopant atoms inside the vdW gap. 
However, neutron scattering experiments (in the case of \CuBiSe{}~\cite{Froehlich2020}) and transmission electron microscopy (TEM) studies (in the case of \SrBiSe{}~\cite{Li2018}) did not observe dopants inside the vdW gap for superconducting single crystals.
In the case of TEM, it was suggested~\cite{Li2018} that lateral movement of the dopants (due to the low diffusion barrier~\cite{Wang2011}) inside the vdW gap at room temperature precludes an observation. However, a recent TEM study of native defects in \BiSe{} demonstrated that Bi atoms inside the vdW gap can be observed~\cite{Callaert2019}.

STM data of Cu-doped \BiSe{} thin-films grown by molecular beam epitaxy showed Cu atoms inside the quintuple layer, but no direct evidence of Cu inside the vdW gap~\cite{Wang2011}.
It was suggested that STM is not sensitive to Cu atoms inside the vdW gap, however, a recent study  of native defects in \BiSe{} by Dai \textit{et al.} \cite{Dai2016} indicated that Se atoms inside the vdW gap can be resolved.
Interestingly, Cu-doped (as well as more recently studied Sr-doped~\cite{Wang2018}) \BiSe{} thin films grown under UHV conditions have carrier concentrations similar to those of bulk single crystals~\cite{Wang2011,Shirasawa2014,Wang2018} but do not show superconductivity.
The absence of superconductivity in these films strongly suggests that, contrary to what was originally proposed \cite{Hor2010}, the role of the dopants in inducing superconductivity in these materials goes beyond simple electron donation into the \BiSe{} lattice.
In this context, it is important to note that Cu- (or Sr-) doped \BiSe{} single crystals only exhibit superconductivity reproducibly when carefully tuned annealing and rapid cooling/quenching are employed~\cite{Wang2016,Shruti2015}.
This indicates that a significant amount of dopants in the superconducting single crystals are trapped in metastable sites (likely not the energetically most favorable sites according to DFT) and that these dopants are indeed responsible for inducing superconductivity \cite{Wang2016,Shruti2015}. 
Thus, deepening the understanding of superconductivity in doped \BiSe{} requires advanced structure characterization in order to elucidate the intricate role played by the dopant. 

\section{Experimental Methods}

Sample preparation and characterization using STM took place in Cologne. 
Single crystals of \SrBiSe{} (nominal $x = 0.06$) were grown from high-purity elemental Sr chunk (99.99\%), Bi shots (99.9999\%), and Se shots (99.9999\%) by a conventional melt-growth method. The raw materials with a total weight of $4.0\,\mathrm{g}$ were mixed and sealed in an evacuated quartz tube. The tube was heated to $850^\circ\mathrm{C}$ for $48\, \mathrm{h}$. It was then slowly cooled from $850^\circ\mathrm{C}$ to $600^\circ\mathrm{C}$ within $80\,\mathrm{h}$ and finally quenched into water at room temperature.

Single crystals of \BiSe{} were grown by melting stoichiometric amounts of Bi and Se shots (99.9999\% purity) in a sealed evacuated quartz tube by the modified Bridgman method. The tube was heated to $850^\circ\mathrm{C}$ for $48\,\mathrm{h}$ with intermittent shaking to ensure the homogeneity of the melt, followed by cooling to $550^\circ\mathrm{C}$ in $100\, \mathrm{h}$. It is then quickly cooled down to room temperature in $2\,\mathrm{h}$.

\begin{figure}[t!]
	\includegraphics[width=8.6cm]{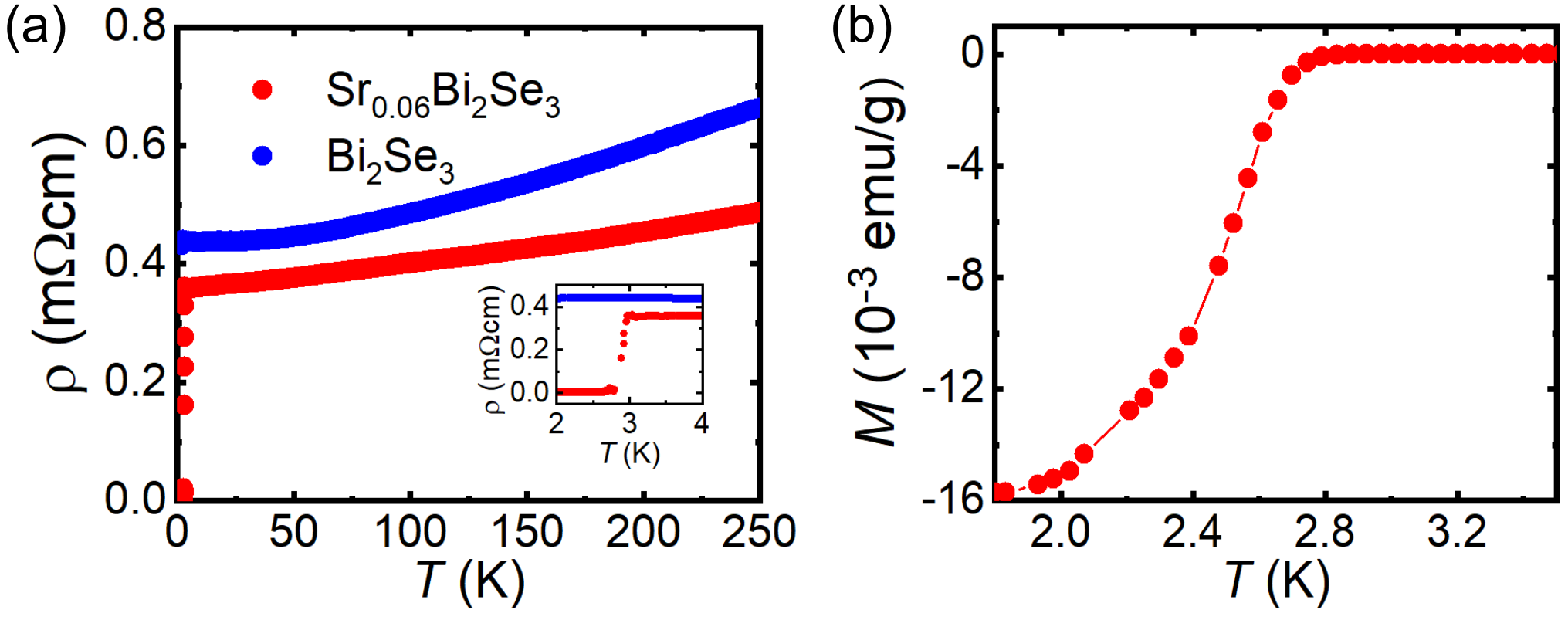}
	\caption{(a) Temperature dependence of the resistivity of \SrBiSe{} and \BiSe. The inset shows the transition to superconductivity of \SrBiSe{} with $T_\mathrm{c}=2.7$~K. (b) Zero-field cooled magnetization curve for ${\mathrm{Sr}}_{x}{\mathrm{Bi}}_{2}{\mathrm{Se}}_{3}$, with a magnetic field of ($B_{\mathrm{DC}}$) of $0.2$~mT applied along the (001) plane. The shielding fraction is nearly 76\% at 1.78 K. }
	\label{resistivityandmag}
\end{figure}

Resistivity measurements on the samples were performed in a quantum design physical property measurement system (PPMS) using the standard four-probe technique. A quantum design superconducting quantum interference device (SQUID) was used to measure the DC magnetization. The results of both measurements are shown in Fig.\ \ref{resistivityandmag}.
The \SrBiSe\ sample was found to be superconducting with a transition temperature of $T_\mathrm{c}=2.7$~K. Its superconducting shielding fraction of $76$\% was estimated from its magnetic moment at $T=1.78$~K, after zero-field cooling.

STM experiments were carried out under ultra high vacuum (UHV) conditions with a commercial system (Unisoku USM1300) operating at $1.7$~K and after cleaving the samples in UHV. Image processing was done with the WSxM software \cite{wsxm}.

Normal incidence x-ray standing waves (NIXSW) as well as photoelectron spectroscopy (PES) experiments were performed at beamline I09 of the Diamond Light Source (DLS), Didcot, UK. The beamline offers the possibility to align both soft and hard x-ray beams onto the same spot on the sample, with a footprint of (in our case) approximately $400 \times 250~\mu$m$^2$. With the soft x-ray beam ($h\nu=700$~eV), we measured standard PES data from all three relevant species using the core levels Se 3d, Bi 5d and Sr 3d. For NIXSW we recorded Se 3s, Bi 5d and Sr 3d spectra at photon energies close to $h\nu=3250$~eV. Prior to these experiments, both samples have been glued, outgassed and transported to the synchrotron under vacuum. At the synchrotron, the \SrBiSe\ crystal was cleaved under UHV conditions. The \BiSe\ crystal, however, was cleaved in argon atmosphere, and then moved to the main chamber using a UHV suitcase. 

With the NIXSW technique \cite{Vartanyants2013,Woodruff2005,Zegenhagen1993}, the vertical positions of atomic species in a crystal or above its surface can be determined with very high precision, usually better than $0.05$~\AA. The technique is based on the interference of an incoming x-ray beam with a Bragg-reflected beam, together forming a standing wave field in the crystal and above its surface, the periodicity of which equals the lattice spacing $d_{hkl}$ of the used Bragg reflection $H=(hkl)$. In one NIXSW scan, the photon energy is tuned through the Bragg condition, i.e., in a narrow energy region around the Bragg energy, under normal incidence of the incoming beam with respect to the $(hkl)$ Bragg planes. In such a scan, the phase difference $\Phi(h\nu)$ between the incoming and the diffracted beams changes from $\pi$ to zero, causing the standing wave field to move through the crystal (perpendicular to the Bragg planes) by half of the lattice spacing. Consequently, the x-ray intensity at every atomic position in the crystal changes as well, which in turn can be monitored by recording the photoelectron yield of the selected atomic species. By doing so, separately for each species that can be distinguished in PES, we can record photoelectron yield profiles that are characteristic for the vertical distance between the atoms and the next underlying Bragg plane. Based on dynamic diffraction theory, the measured yield curves can be fitted using 
\begin{align} 
\label{eq:yield}
Y(h\nu) = 1 & + S_R R(h\nu) \\
         ~  & + 2 \left| S_I \right|  \Fc \sqrt{R(h\nu)} \cos{ \left(\Phi(h\nu) - 2\pi \Pc + \psi \right)},  \nonumber
\end{align} 
with the Bragg-reflected x-ray intensity $R(h\nu)$, the correction factors for non-dipolar effects $S_R$ and $S_I = \left| S_I \right| e^{i \psi}$, and the parameters coherent fraction $\Fc$ and coherent position $\Pc$ \cite{Bocquet2019,Woodruff2005a,vanStraaten2018}. 
The latter two, both spanning from 0 to 1, are obtained from a fit of Eq.\ \ref{eq:yield} to the experimental data. These quantities represent the structural information that can be obtained from a NIXSW experiment. $\Pc$ is the averaged vertical distance of the selected atomic species to the nearest Bragg plane underneath, in units of the lattice spacing $d_{hkl}$. $\Fc$ contains information on the vertical order of the atoms, including the occurrence of several well-defined atomic sites (heights) in the crystal, as it will be relevant in this work. A value of $1.0$ usually indicates that all atoms occupy sites that have the same distance to the next Bragg plane, and that they are (vertically) perfectly ordered. The fact that NIXSW is very sensitive to vertical positions, but does not rely on lateral order, makes the method well-suited for the investigation of layered 2D materials, where positions may be expected to be constrained more stringently in the vertical direction. The specific cases of \BiSe\ and \SrBiSe\ samples will be discussed in detail in section \ref{sec:Results-XSW}. 

\section{Experimental Results}

The general goal of our work is to identify the (vertical) positions of the Sr-dopants in the \SrBiSe\ crystal, since these are of utmost importance for understanding the superconductivity of this material.
Motivated by previous STM studies on near-surface defects in \BiSe{}~\cite{Dai2016, Urazhdin2002}, we performed similar measurements on our \BiSe{} and \SrBiSe{} crystals, hoping to gain insights into the position of the Sr dopants by comparing the results of doped and undoped samples (see following section). In addition to this local technique, we also applied the integrating NIXSW technique (see Sec.\ \ref{sec:Results-XSW}), which turns out to be ideally suited for our problem and allowed us to unambiguously identify the dominant vertical position of the dopant species.

\subsection{Near-surface defects seen in STM}
\label{sec:Results-STM}

\begin{figure}[t!]
	\includegraphics[width=8cm]{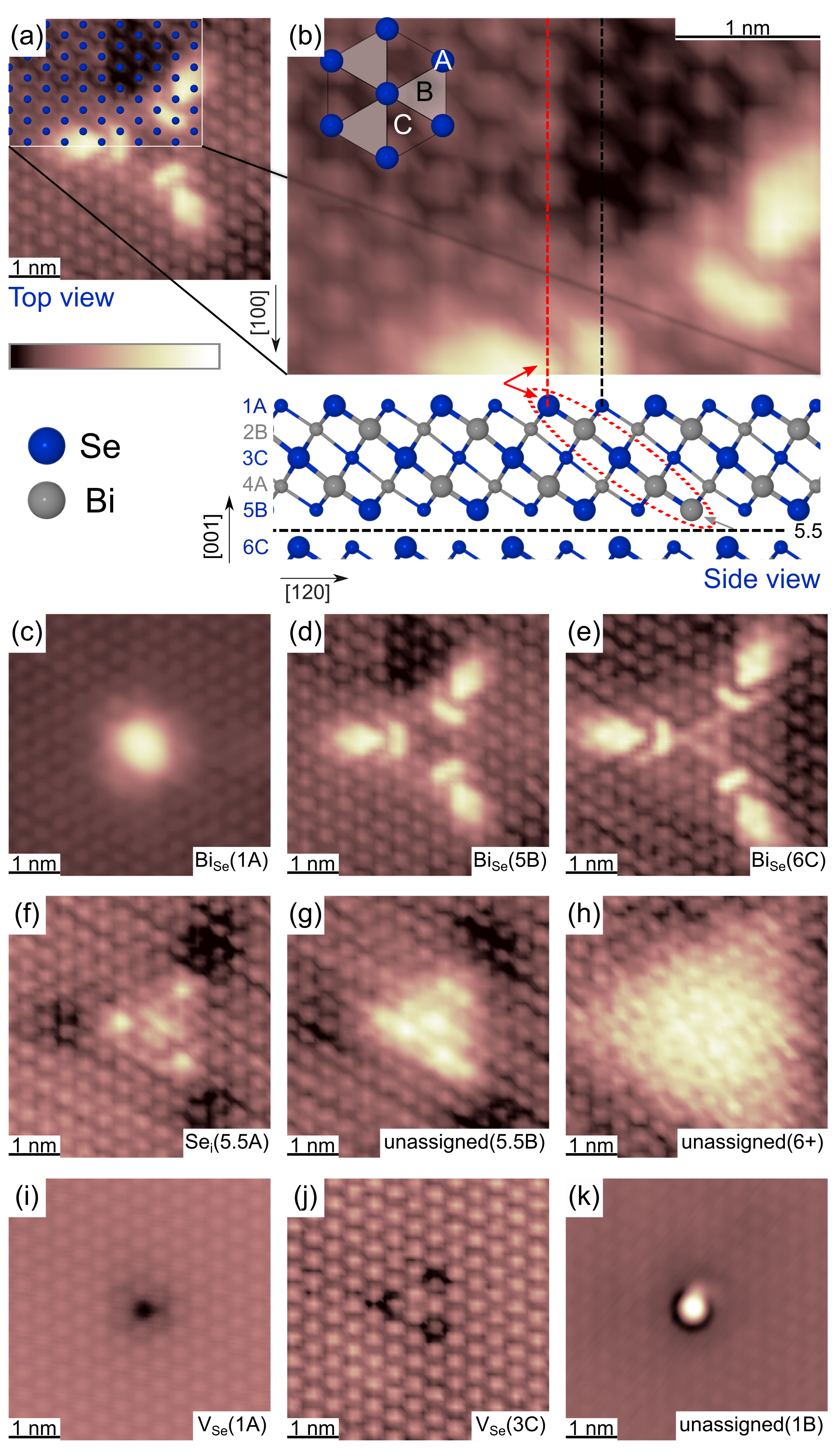}
	\caption{Typical STM images of all relevant defects found both in \SrBiSe{} and in \BiSe. (a) Bi$_\text{Se}$(5B) defect. In a part of the image the Se1 sites are marked by blue spheres. (b) Top view (section of (a)) and side view (model) of the Bi$_2$Se$_3$ structure. The red dashed ellipse marks the Se-Bi-Se-Bi-Bi$_\text{Se}$ chain, along which the antisite in the 5B position modifies the LDOS around the 1A Se atom (red arrow). A, B, and C sites are defined in the inset. (c)-(e) Bi antisites Bi$_\text{Se}$ at 1A, 5B, and 6C positions, respectively. (f) Se interstitial Se$_\text{i}$ between layer 5 and 6 in site (5.5A). (g),(h) Unassigned defects in site 5.5B and layer 6 or below. (i),(j) Se vacancies V$_\text{Se}$ at 1A and 3C. (k) Unassigned surface defect in site (1B). Image parameters: (a)-(k) $I_T=100~\text{pA}$, (c),(h) $U_T=-0.6$~V, (d)-(g),(i),(j) $U_T=-0.9$~V, and (k) $U_T=-0.7$~V. 
	The color scale for the images (a)-(k) are 30~pm, 30~pm, 300~pm, 30~pm, 40~pm, 25~pm, 30~pm, 10~pm, 160~pm, 25~pm, 95~pm, respectively. All images were recorded at $T=1.7$~K.}
	\label{fig:NativeDefectSTMSummary}
\end{figure}

First, we briefly recall how subsurface defects in \BiSe{} are imaged by STM.
A simple but instructive picture was introduced by Urazhdin \textit{et al.}~\cite{Urazhdin2002}: 
The local density of states (LDOS) of valence and conduction bands of \BiSe{} is formed exclusively from p-orbitals of Se and Bi, respectively.
Moreover, the bonding within a quintuple layer can approximately be viewed as strongly interacting atomic p-orbitals ($pp\sigma$) along Se-Bi-Se-Bi-Se chains. 
Therefore, a subsurface defect will lead to a perturbation of the p-orbitals along the chain affecting the LDOS around the surface Se atom.
This mechanism is readily seen in the extensively studied Bi-antisites \cite{Urazhdin2002,Dai2016}. For example, a Bi$_\text{Se}$(5B) antisite, that is a Bi atom replacing a Se atom in the fifth atomic plane beneath the surface, causes an increase in the LDOS at the surface as shown in Figs.\  \ref{fig:NativeDefectSTMSummary}(a), (b), and (d).
Note that three symmetry-equivalent chains, rotated by $120^\circ$ with respect to each other around the [001] direction, give rise to the triangular patterns observed in STM images (\fig{fig:NativeDefectSTMSummary}~(d)). 
Within this simple picture, we can use the lateral size of the triangular pattern visible in the top Se layer to infer the depth of the Bi-antisite: 
Bi$_\text{Se}$(6C) (\fig{fig:NativeDefectSTMSummary}~(e)) appears as a larger triangle than Bi$_\text{Se}$(5B) (\fig{fig:NativeDefectSTMSummary}~(d)),  and Bi$_\text{Se}$(1A) appears almost point-like (\fig{fig:NativeDefectSTMSummary}~(c)).  

\begin{figure}[t!]
	\includegraphics[width=8.6cm]{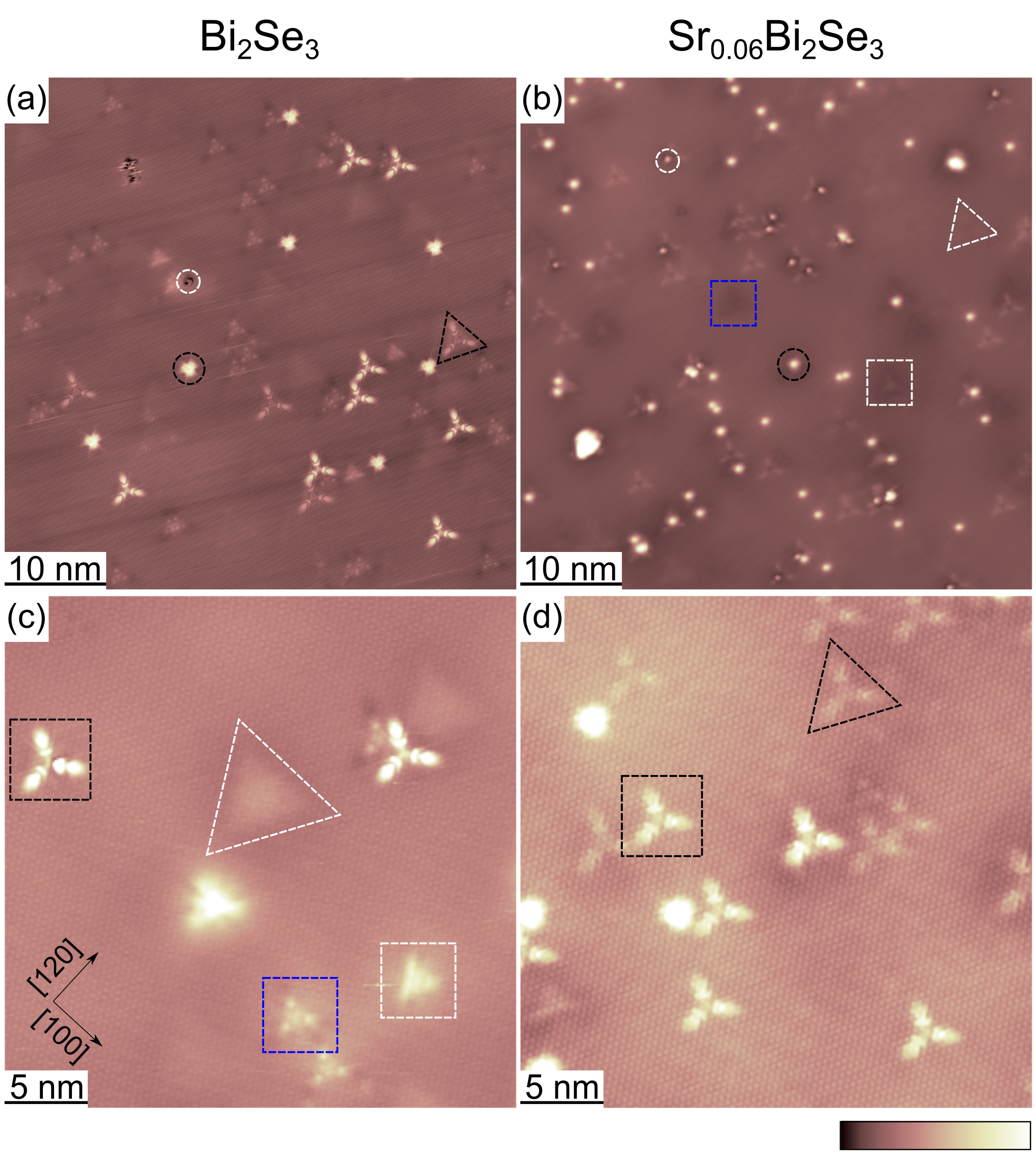}
	\caption{{Typical STM images of the (001) plane of (a),(c) Bi$_2$Se$_3$ and (b),(d) Sr-doped Bi$_2$Se$_3$ single crystals directly after cleaving. The same defects are visible in both crystals: Bi$_{\text{Se}}$(1A)/(5B)/(6C) (black circle/square/triangle), unassigned (1B)/(5.5B)/(6) (white circle/square/triangle), and  Se$_{i}$(5.5A) (blue square). Tunneling parameters: (a)  $U=-900$~mV, $I=20$~pA; (b) $U=-900$~mV, $I=0.2$~nA; (c) $U=-600$~mV, $I=5$~nA; (d) $U=-650$~mV, $I=0.5$~nA. Color scale of the STM images: (a),(b) 250~pm, (c) 50~pm, and (d) 100~pm.}}
	\label{fig:STMComparisonBiSeSrBiSe}
\end{figure}

This intuitive picture was corroborated in the comprehensive study by Dai \textit{et al.}~\cite{Dai2016}, who identified various native defects in \BiSe{} by comparison of STM data with simulated images based on first-principles calculations.
From here on, we follow the nomenclature introduced in Ref.~\cite{Dai2016} and classify the defects observed in our crystals as Bi-antisites (Bi$_\text{Se}$), Se-vacancies (V$_\text{Se}$), and Se-interstitial (Se$_\text{i}$), based on their characteristic triangular shapes in high resolution STM images (Fig.~\ref{fig:NativeDefectSTMSummary}). 
Moreover, we also observe several defects which were not previously reported and denote them as unassigned.
Se$_\text{i}$(5.5A), see Fig.~\ref{fig:NativeDefectSTMSummary}~(f), and the defect shown in Fig.~\ref{fig:NativeDefectSTMSummary}~(g) have a similar lateral size, but the latter is centered around site B (see the inset of Fig.~\ref{fig:NativeDefectSTMSummary}~(b) for the definition of A, B, and C-sites).  
The defect in \fig{fig:NativeDefectSTMSummary}~(h) appears laterally similarly extended as Bi$_\text{Se}$(6C) and is therefore assigned to a defect in the Se6 layer.
The almost point like defect in \fig{fig:NativeDefectSTMSummary}~(k) is straightforwardly assigned to the 1B site.

A side-by-side comparison of typical larger scale STM images of \BiSe{} and \SrBiSe{} surfaces (\fig{fig:STMComparisonBiSeSrBiSe}) shows all defects discussed above for both crystals: none is uniquely found in \SrBiSe{}. Since it is conceivable that the Sr dopants have a similar appearance in STM images as (some) of the native defects shown in \fig{fig:NativeDefectSTMSummary} and may be mistaken for these other defects, we proceeded to quantify the relative abundance of different defects in our \BiSe{} and \SrBiSe{} crystals to gain further insights into possible Sr dopant locations.
The defect densities in both crystals were deduced by evaluating several STM images covering a total surface area of about 77000 (37000)~$\mathrm{nm}^2$ for \BiSe{} (\SrBiSe{}) and the results are summarized in Table~\ref{tab:TableDefectDensitiy}. The defect density of V$_{\text{Se}}$ is much lower than that of other types of defects.
Moreover, the V$_{\text{Se}}$ defect density in \SrBiSe{} is only half of that in \BiSe{}. Therefore we disregard that Sr dopants can be mistaken for V$_{\text{Se}}$.
While the density of Se$_{\text{i}}$ is significantly higher than for V$_{\text{Se}}$, there is almost no difference between \BiSe{} and \SrBiSe{}, which makes it unlikely that (some) of the Se interstitial defects are in fact due to Sr dopants.
This is different for the defect densities of the Bi$_{\text{Se}}$ and unassigned defects in \SrBiSe{}, which are about 10 and 2 times higher compared to \BiSe{}, respectively.
However, all of the observed defect densities resolved by STM are more than an order of magnitude lower than the  density of Sr dopants (about $420\times10^{18}\, \mathrm{cm^{-3}}$)  expected from the nominal doping level ($x=0.06$).

\begin{table}
  \begin{tabular}{lccccc}
    \hline\hline
  Sample & Bi$_{\text{Se}}$ & Se$_{\text{i}}$ &  & V$_{\text{Se}}$ &  unassigned \\
	\hline
  Bi$_2$Se$_3$ & $1.8$ & $3.0$ & & $0.8$ & $3.0$  \\
  Sr$_x$Bi$_2$Se$_3$ & $17$ & $2.3$ &  & $0.4$ & $6.5$   \\
  \hline\hline
\end{tabular}
\caption{Defect density (in units of $10^{18} \mathrm{cm}^{-3}$) of bismuth antisites (Bi$_\text{Se}$), selenium interstitial (Se$_\text{i}$), Se vacancies (V$_\text{Se}$) and unassigned defects observed by STM in the top quintuple layer.
\label{tab:TableDefectDensitiy}
}
\end{table}

In summary, the comparison of \BiSe{} and \SrBiSe{} surfaces by STM showed the same defects, which were classified into Bi-antisites (Bi$_\text{Se}$), Se-interstitials (Se$_\text{int}$), and Se vacancies (V$_\text{Se}$) according to their characteristic appearance in high resolution STM images (\fig{fig:NativeDefectSTMSummary}) and by comparison with the work by Dai \textit{et al.}~\cite{Dai2016}.
While some defects in our crystals could not be assigned by this comparison, they were seen in both doped as well as undoped \BiSe{}. 
Therefore, no straightforward identification or Sr dopants based on their characteristic appearance in STM images was possible.
The quantitative comparison of defects in both crystals revealed an increase in both Bi-antisites (Bi$_\text{Se}$) and unassigned defects, both of which are situated in or near the Se1 and Se5 layer.
However, even when assuming that the increase in both defect types is in fact due to Sr-dopants, it would only account for about 4\% of the Sr-dopants expected in our \SrBiSe{} crystals.

The inconclusive results from our local probe measurements lead us to perform element specific and surface averaging NIXSW experiments which are discussed in the following.

\subsection{Vertical structure revealed by NIXSW}
\label{sec:Results-XSW}

\subsubsection{Structure model}

\begin{figure}
	\centering
	\includegraphics[width=\columnwidth]{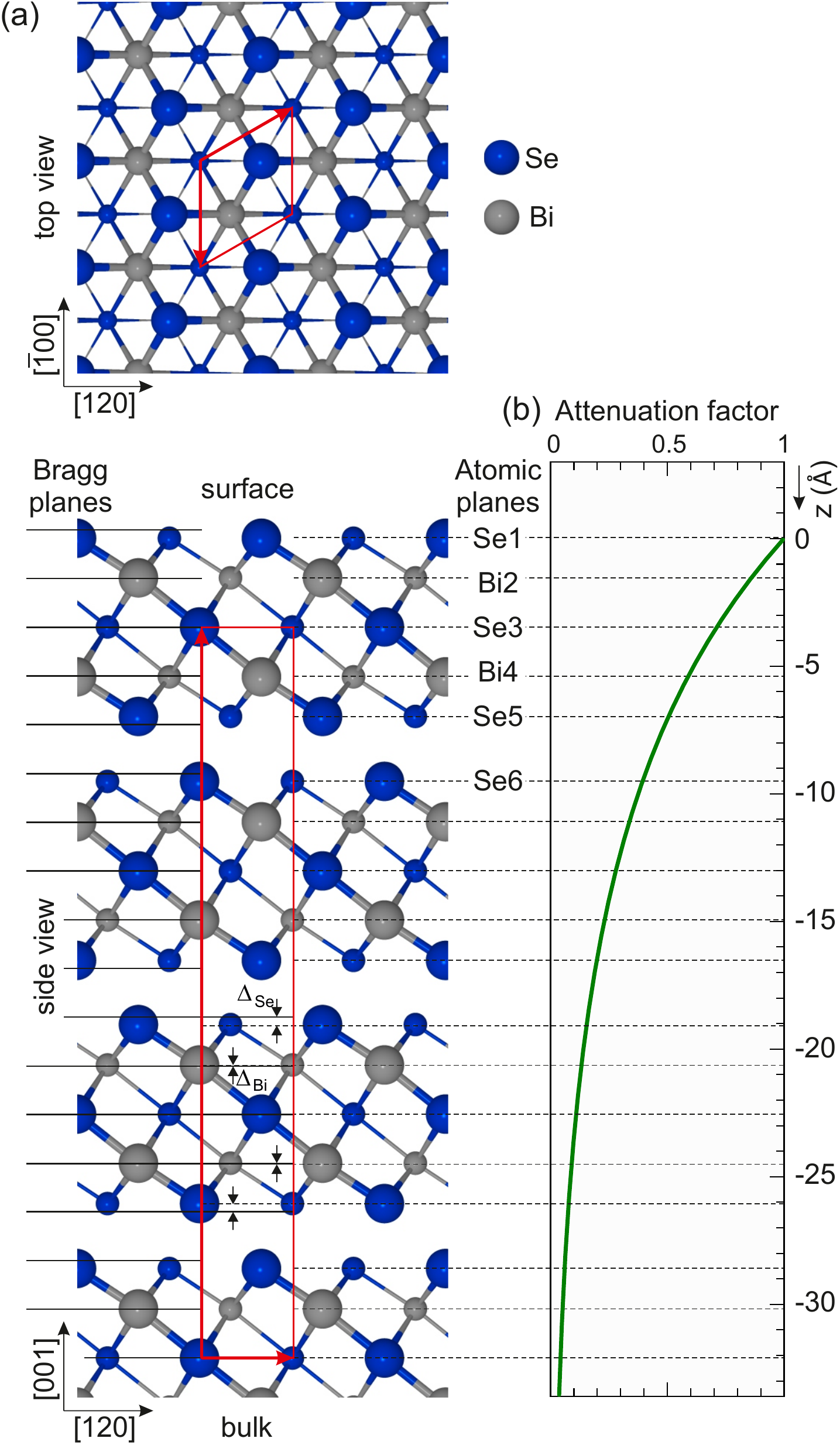}
	\caption{\label{fig:Model}
		(a) Ball-and-stick model of the \BiSe{} bulk crystal, shown as top view along the $[00\overline{1}]$-direction (upper part) and side view along $[\overline{1}00]$ (lower part). The sizes of balls and sticks are linearly scaled with the distance of the atoms to the front plane (in the top view that is the crystal surface, in the side view it is the plane spanned by $[120]$ and $[001]$ corresponding to the lower edge of the top view). The unit cell (in the side view its projection) is indicated by the red parallelogram and rectangle. In the side view we also indicate the $z$-positions of the atomic planes (right part, dashed lines extending into (b)) and the $(0~0~0~15)$ Bragg planes (left part, solid lines). (b) Attenuation factor for the yield of a photoemission process occurring in the depth $z$, according to a continuous exponential damping with an inelastic mean free path of $\lambda = 65~$\AA, as obtained from fitting the NIXSW data, and an emission angle with respect to the surface normal of $\phi = 80.9^\circ$. }
\end{figure}

The structure model that we used for analyzing our data is shown in Fig.~\ref{fig:Model}. Panel (a) shows the \BiSe{} crystal structure in top view (upper part) and side view (lower part). The top view contains the atoms of the uppermost three atomic layers only, drawn as blue (Se) and gray (Bi) balls, the size of which indicates the depth below the surface. In the side view, we show the crystal terminated by a complete quintuple layer (QL, atomic layers from Se1 to Se5). The hexagonal unit cell is indicated by red lines and contains three QLs, i.e., 15 atomic planes. The vertical ($z$-) positions of the atomic planes are marked as dashed lines in the right half of the figure (extending into Fig.~\ref{fig:Model}(b)). For NIXSW we used the $H=(0~0~15)$ Bragg reflection ( $(hkil)=(0~0~0~15)$ in hexagonal notation), and thus we find the same number of Bragg planes, 15, within the unit cell. They are equidistantly spaced in $z$, in contrast to the atomic planes. Therefore, while the central Se layer of each QL (Se3, Se8, ...) lies precisely on a Bragg plane, all other atomic planes lie close to but not on the Bragg planes, as indicated by the small gaps labeled $\Delta_\text{Se}$ and $\Delta_\text{Bi}$ in the side view of Fig.~\ref{fig:Model}(a). This will be important for the interpretation of the NIXSW data (see below). 

\subsubsection{NIXSW data}

\begin{figure}
  \centering
  \includegraphics[width=\columnwidth]{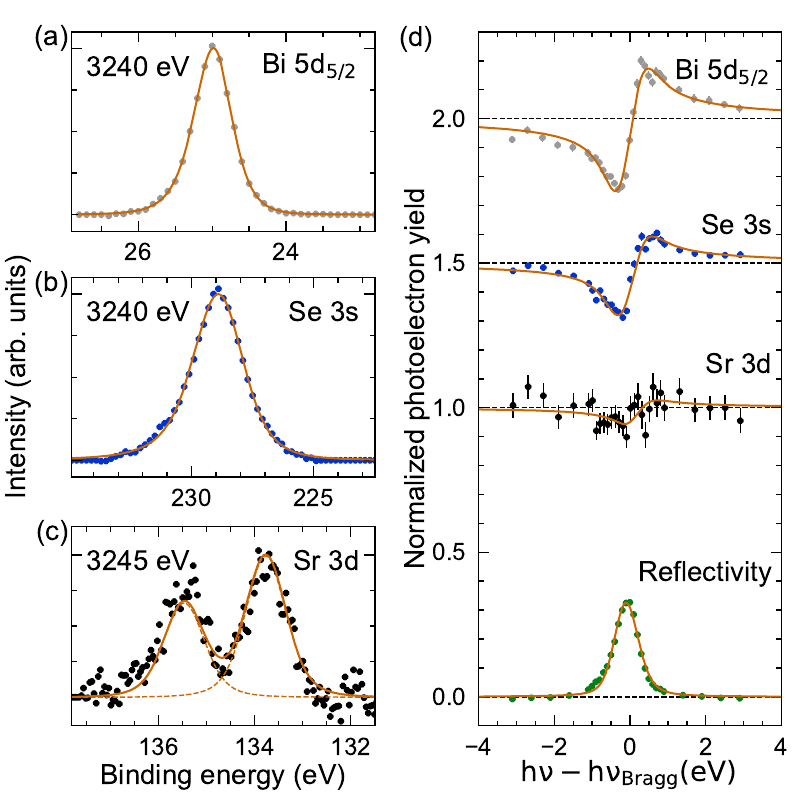}
  \caption{\label{fig:XPS-yield}
  NIXSW results for the Sr-doped \BiSe{} sample. (a-c) Typical core level spectra of Bi 5d$_{5/2}$, Se 3s and Sr 3d, respectively. The data were recorded about $3$~eV off the Bragg energy. (d) NIXSW photoelectron yield curves (top), obtained from fitting XPS data that were recorded in an energy window of $\pm 3$~eV around the Bragg energy of the \BiSe\ $(0~0~0~15)$ reflection, and reflectivity curve of this reflection (bottom). For Bi and Se individual XPS and NIXSW scans are displayed, while for Sr the sum of five individual scans is shown because of the low count rate, see also Sec.\ \ref{sec:Appendix} (Appendix). The curves for Se and Bi are displaced vertically by 0.5 and 1.0, respectively.}
\end{figure}

\begin{figure}
	\centering
	\includegraphics[width=\columnwidth]{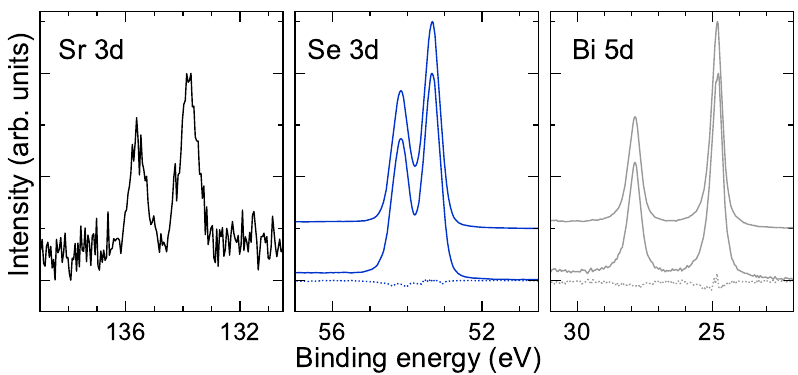}
	\caption{\label{fig:soft-XPS} Core level spectra for \BiSe{} and \SrBiSe{}, taken at photon energy of $700$~eV. The Se 3d and Bi 5d spectra of the undoped (upper) and doped (lower) \BiSe{} sample are vertically displaced in order to improve visibility. Difference plots (\SrBiSe{}$-$\BiSe{} ) are shown as dotted lines.}
\end{figure}

We have recorded NIXSW data sets for an undoped and a Sr-doped \BiSe{} sample. Bi 5d, Se 3s and (in case of the doped sample) Sr 3d core level spectra have been recorded in a $6$~eV interval around the Bragg energy $h\nu_\mathrm{Bragg}= 3256$~eV of the $(0~0~0~15)$ Bragg reflection, and the partial photoelectron yield was extracted for all species as a function of the photon energy. We fitted the spectra with \textsc{CasaXPS} \cite{CasaXPS} using one asymmetric Voigt peak for Se 3s and Bi 5d$_{5/2}$, and two symmetric Voigt peaks for the Sr 3d doublet with their relative intensity constrained to a ratio of 2:3. Exemplary spectra and fit curves (for \SrBiSe{} only) are shown in Fig.~\ref{fig:XPS-yield}(a)-(c). These simple fitting models were sufficient, since no indications for multiple components in different chemical environments were detected in XPS, neither with the hard x-ray beam, nor using the soft x-ray beam that allows for a higher resolution. The latter data are shown in Fig.~\ref{fig:soft-XPS} and demonstrate that no significant spectroscopic differences were detected between the \BiSe{} and the \SrBiSe{} samples. This indicates that -- as expected from the small dopant concentration --  the dopants do not cause any spectroscopically relevant change to the chemical environment of the bulk species. 
Taking into account the photoemission cross-sections at the x-ray energies used, the relative intensities of the Sr, Bi and Se PES peaks are in qualitative agreement with the expected stoichiometry of \SrBiSe{}.

Typical partial yield curves, together with the x-ray reflectivity, are shown in Fig.~\ref{fig:XPS-yield}(d), again only for the \SrBiSe{} data set. The yield curves were fitted using our dedicated NIXSW analysis software \textsc{Torricelli} \cite{Bocquet2019, Torricelli}, which is able to consider non-dipolar effects in the photoemission process as well as a finite deviation from the ideal normal incidence geometry that cannot be avoided in the experiment \cite{NDP, vanStraaten2018}. All final results (after averaging of up to 7 equivalent measurements) are listed in Table \ref{table:XSWresults}. Despite the statistics of the Sr yield curve being apparently poor, obviously caused by the small density of Sr atoms in the crystal, the Sr data actually imposes a sufficient constraint on the fitting for reliably extracting values for the Sr coherent position and fraction, as demonstrated in Sec.\ \ref{sec:Appendix} (Appendix).

We note that the experimentally obtained coherent positions $\Pc$ for all species of both samples \BiSe{} and \SrBiSe{} are not zero. For Se and Bi this is remarkable, since the symmetry of the layer structure of the QLs (and the entire unit cell) should actually cancel out any effect that causes a finite coherent position, as discussed in detail in the following section.

\subsubsection{The effect of a finite inelastic mean free path}

Commonly, the NIXSW technique is used as a method to determine the height of specific atomic species above a surface, which is very useful to evaluate, e.g., the interaction of adsorbates (atoms or molecules) with the surface or with other adsorbates on the surface \cite{Woodruff2005, Woolley2007, Duhm2013, Stadtmueller2014, Stadtmueller2015, Goiri2014, Mercurio2014, Baby2017, Stadtmueller2016, Blowey2019, Klein2019, Bocquet2020}.
In such experiments, NIXSW data of the bulk species are often also recorded, in order to monitor the surface quality or simply for reasons of completeness. For crystals with a mono-atomic primitive unit cell, the structural parameters obtained by NIXSW for the bulk species are very close to $\Pc=0.0$ and $\Fc=1.0$ in many cases, indicating that the atoms are located on Bragg planes (which is the case per definition) and that the crystal is well ordered. But in multinary crystals both coherent fraction and coherent position may vary from these values, depending on the number of inequivalent layers of the relevant species and their position(s) relative to the Bragg planes. 
When bulk species are considered, attenuation effects due to the finite inelastic mean free path have to be considered additionally \cite{Paez2020}, as discussed below.

\begin{figure}
  \centering
  \includegraphics{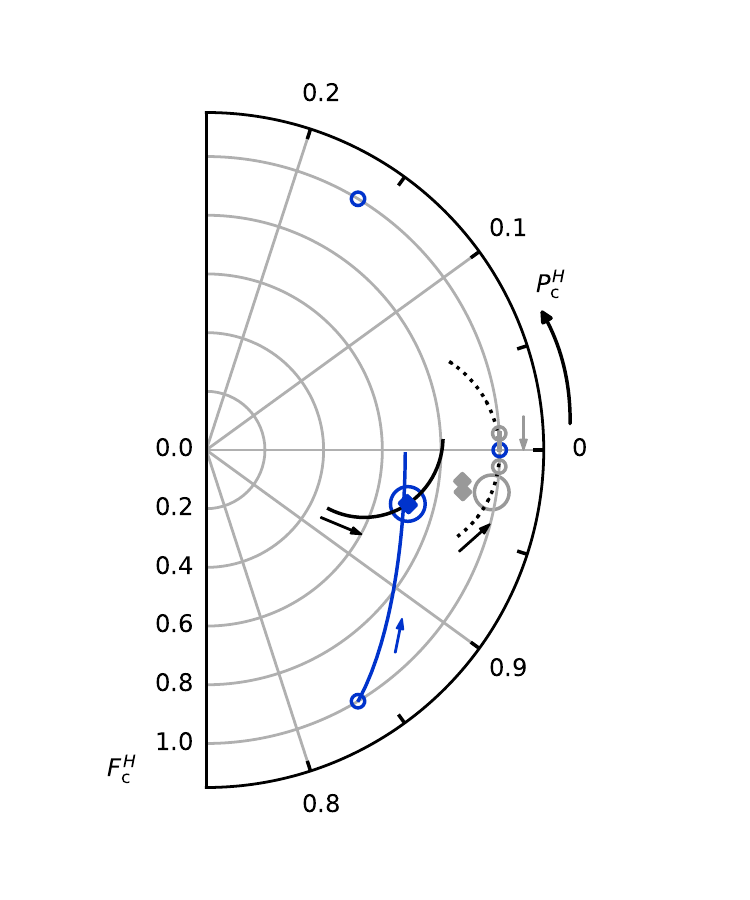}
  \caption{     \label{fig:Argand}
Illustration of the NIXSW results for \BiSe{} in an Argand diagram. The data points represent radial vectors (Argand vectors), the lengths and polar angles of which correspond to the coherent fractions and positions, respectively. Experimental data (all individual NIXSW scans) obtained for PE from Se 3s (Bi 5d$_{5/2}$) are shown as blue (gray) diamonds, numbers listed in Table \ref{table:XSWresults} represent their average. Small open circles (three for Se, two for Bi) indicate the Argand vectors calculated for each individual atomic plane within one QL. Averaging the contributions of the individual layers, under variation of the inelastic mean free path $\lambda$ (from zero to infinity in the direction of the small arrows), results in Argand vectors ending on the blue (gray) curved line for Se (Bi). Note that the gray curved line is very short, located close to $(\Fc, \Pc) = (1.0, 0.0)$. Black lines (solid and dotted) indicate the variation of the expected NIXSW results for a relaxation of the uppermost layers of the Se and Bi species, respectively. They are calculated for $\lambda = 65$~\AA, and a shift of the Se1 and Bi2 layer from $-0.3$ to $+0.3$~\AA\ (also in the direction of the small black arrows). The best agreement with the measured data is obtained for an Se1-Bi2 expansion of $9.7$\% to $\Delta_\text{Se1-Bi2} = 1.73$~\AA, as indicated by large open circles.
   }
\end{figure}

In our case, for the bulk species of both \BiSe{} and \SrBiSe{}, there are two (for Bi) and three (for Se) inequivalent layers within the unit cell, namely those within one QL. This is illustrated in the side view model shown in Fig.~\ref{fig:Model}(a). All layers, except the central Se3 layer of a QL, are not located on diffraction planes of the ($0~0~0~15$) Bragg reflection that we used for the NIXSW measurements. The difference is very small for the Bi planes ($\Delta_\text{Bi} = 0.02~$\AA), but much more significant for Se ($\Delta_\text{Se} = 0.31~$\AA), and has different signs (negative for Se1 and Bi2, positive for Bi4 and Se5). Consequently, the coherent positions of the individual atomic planes are equal to zero only for the central Se3 layer, not for the other four layers. For Se1 and Se5 one expects $P_\mathrm{c, Se1,5}^{H} = \pm 0.164$, for Bi2 and Bi4 $P_\mathrm{c, Bi2,4}^{H} = \pm 0.009$. This is illustrated in the Argand diagram presented in Fig.~\ref{fig:Argand}. Data points in this diagram represent radial vectors with their length corresponding to $\Fc$ and their polar angle to $\Pc$. Multiple-site emission, as it occurs in the \BiSe{} crystal, is then accounted for by the vector average of the Argand vectors representing the individual layers of identical species. For the three Se layers Se1, Se3 and Se5 of one QL, the individual Argand vectors are indicated by small blue circles at $(\Fc, \Pc) = (1.0, 0.164)$, $(1.0, 0.0)$, and $(1.0, 0.836)$ in Fig.~\ref{fig:Argand}. Note that the latter is equivalent to $(1.0, -0.164)$, and hence the average of the three vectors, which corresponds to the expected experimental result of a Se based NIXSW measurement, would be $(0.678, 0.0)$, i.e., in particular, the coherent position would be zero for such a measurement. This is to be expected, as long as the mirror symmetry of the full QL with respect to the central Se layer is in place.

\begin{table}[t]
	\caption{\label{table:XSWresults} 
		NIXSW results for \BiSe{} and \SrBiSe{}, as averaged from all individual NIXSW scans. The data are corrected for non-dipolar effects and a deviation from normal incidence geometry ($2\theta < 180^\circ$) \cite{NDP, vanStraaten2018}. 
	}
	\begin{tabular}{p{16mm}p{12mm}p{12mm}p{3mm}p{12mm}p{12mm}}
		\hline \hline
		& \multicolumn{2}{c}{\BiSe{} }            & & \multicolumn{2}{c}{\SrBiSe{} }                  \\
		& \centering $\Pc$  & \centering $\Fc$ & & \centering $\Pc$ & \multicolumn{1}{c}{$\Fc$} \\  
		\hline
		Bi 5d$_{5/2}$  & 0.98(2)       & 0.89(1)          & & 0.95(2)         & 0.74(2)   \\
		Se 3s          & 0.96(2)       & 0.71(1)          & & 0.91(2)         & 0.64(1)  \\
		Sr 3d          & \centering -- & \centering --    & & 0.86(2)         & 0.49(7) \\
		\hline \hline
	\end{tabular}
\end{table}

However, this symmetry is broken in our NIXSW experiment, not for structural reasons, but in its effect on the photoelectron yield that is obtained from the individual layers, due to the finite inelastic mean free path $\lambda$ of electrons in matter. The photoemission yield that can be recorded from deeper layers is attenuated by a factor of $\alpha(z) = \exp(z/(\cos\phi\cdot\lambda))$, with $\phi = 80.9^\circ$ being the angle between the surface normal and the direction towards the electron analyzer. (Note that we defined the positive $z$ axis towards the vacuum, and hence $z<0$ for all atomic layers below the surface layer Se1.) Hence, all layers of the structure -- even if they are structurally equivalent -- contribute differently to the total photoelectron yield that is experimentally accessible. The attenuation factor, as found by fitting our experimental data (see below), is plotted in Fig.~\ref{fig:Model}(b). 

The attenuation effect has obvious consequences for the averaging of Argand vectors. In the example mentioned above, the Argand vectors representing the three Se layers Se1, Se3 and Se5 have to be scaled in their lengths before averaging (i.e., their effective coherent fractions become $\alpha(z) \Fc$) and hence depend on the depth $z$ of the corresponding layers below the surface. Accordingly, the averaged coherent position $\Pc$ will deviate from zero in our example, since Se1, the layer with $\Pc = -0.164$, is dominant due to its higher scaling factor. Depending on the specific value for $\lambda$, the resulting averaged Argand vector ends on the blue line shown in Fig.~\ref{fig:Argand}, between the two extreme cases of $(\Fc, \Pc) = (0.678, 0.0)$ for $\lambda \rightarrow \infty$ (all three layers contribute equally) and $(\Fc, \Pc) = (1.0, 0.836)$ for $\lambda \rightarrow 0$ (only the uppermost layer contributes). It can be seen that the experimental data points representing the Se 3s NIXSW results (blue diamonds) lie very close to the blue curve, which allows us to estimate the inelastic mean free path in the undoped \BiSe. Note that we performed the same calculation also for Bi2 and Bi4, as shown in Fig.~\ref{fig:Argand} with small gray circles and lines. However, the effect is much smaller for Bi due to the small distance $\Delta_\mathrm{Bi}$ between the corresponding atomic and Bragg planes, and the experimental data points (gray diamonds) lie further away from the calculated ones (gray line running from $(\Fc, \Pc) = (1.0, 0.009)$ to $(1.0, 0.000)$ for $\lambda = 0$ to $\infty$).

With what has been said so far, it is clear that the inelastic mean free path effect fully explains our experimental NIXSW results for Se 3s, but is not sufficient to account for those of Bi, regarding both coherent fraction and position ($(\Fc, \Pc) = (0.98, 0.89)$). For $\Fc$ this is not surprising, since the data was taken using a d-state emission line (the Bi 5d$_{5/2}$), and hence the correction of non-dipolar effects cannot be properly performed for this species \cite{NDP}. But for the coherent position $\Pc$, usually much less affected by non-dipolar effects, the situation can only be improved when a relaxation of the uppermost layers of the crystal is taken into account. Any outward (inward) relaxation of an atomic layer increases (decreases) the coherent position $\Pc$ of that layer, and hence rotates its Argand vector counterclockwise (clockwise). We have analyzed a number of reasonable scenarios for layer relaxations, e.g., a rigid shift of the entire first QL or a relaxation of the uppermost Bi and Se layers. We found that a relaxation of the uppermost layers Se1 and Bi2 gives the best result. Fig.~\ref{fig:Argand} illustrates the effect of Se1/Bi2-relaxation for the averaged NIXSW yield as curved black lines. The solid black line represents the resulting variation of the averaged Argand vector for values of the Se1 relaxation between $-0.3$~\AA\ and $+0.3$~\AA\ (in the direction of the small black arrow), calculated for $\lambda = 65$~\AA. The dotted black line represents the result of a similar model calculation for Bi. It can be seen that this additional parameter helps significantly in explaining our experimental results, in particular for the coherent position of Bi. 

\begin{figure*}[t]
	\centering
	\includegraphics[width=0.8\textwidth]{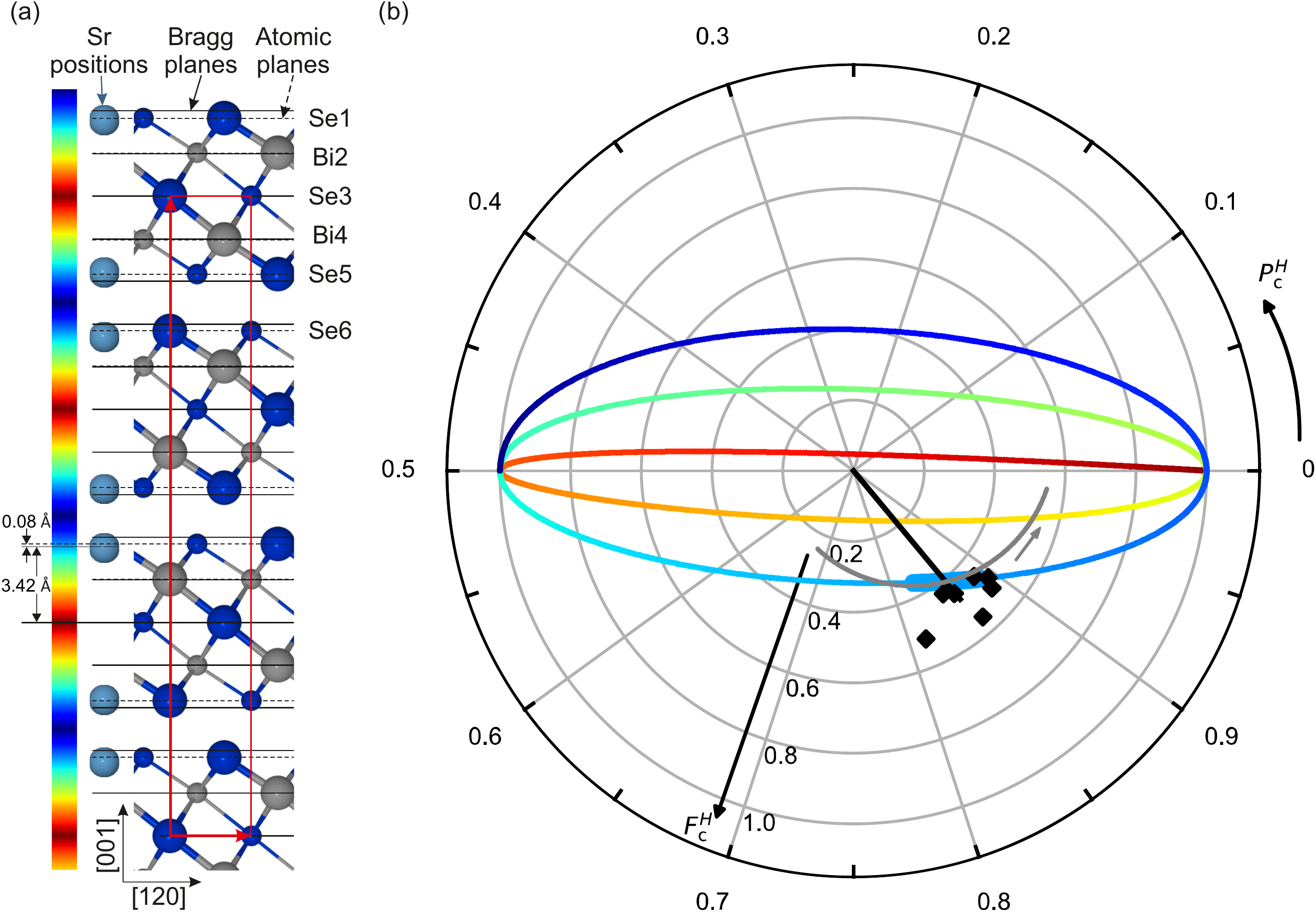}
	\caption{     \label{fig:Sr-Argand}
		Simulation of all possible Sr sites. (a) Atomic model (side view) of the \BiSe{} structure, see also Fig.~\ref{fig:Model}. The color bar in the left illustrates all possible (vertical) Sr dopant positions, whereby symmetry equivalent positions are marked in the same color. The positions vary from the center of a vdW gap (dark blue) to the Se3 layer in the center of a QL (dark red). (b) Rainbow-colored spiral-type line: Argand representation of coherent positions and fractions as calculated for every possible Sr site, considering the unit cell symmetry. Each point on the line represents the head of the Argand vector $(\Fc, \Pc)$, its color corresponds to the vertical atomic position as shown by the color bar in the left of panel (a). For more details see text. Black data points and black radial line: Results of seven individual Sr3d NIXSW measurements and their average, respectively. The experimental data can only be reasonably explained by Sr positions represented by the cyan line section; the region of good agreement of calculated and experimental results is indicated by a thick cyan line. In the left of panel (a), the corresponding atomic positions in the unit cell are indicated by cyan balls.
	}
\end{figure*}

Finally, we optimized all parameters (that is the inelastic mean free path and the relaxation parameters for Se1 and Bi2) together in a least squares fit on the coherent positions of both Bi and Se, and the coherent fraction of Se \cite{NDP}. This resulted in $\lambda = 65$~\AA~and $\Delta_\text{Se1-Bi2} = 1.73$~\AA, the latter corresponding to a $9.7$\% expansion of this interlayer spacing. First, the fitted value of $\lambda$ is in good agreement with calculated values of the electron inelastic mean free path, which show little variation between different materials at high kinetic energies (55~\AA\ to 80~\AA\ for a kinetic energy of about 3.1~keV \cite{Shinotsuka2019}).
Second, we would like to mention that, although NIXSW is not the best method to determine surface relaxation effects, our results are in qualitative agreement with Roy \etal, who found a relaxation of up to  $11$\% for the Se1-Bi2 interlayer distance \cite{Roy2014}.

\subsubsection{Strontium sites in \SrBiSe}

We now turn to a discussion of the NIXSW results on the Sr-doped crystal, with the goal of identifying the (vertical) positions of the Sr atoms in the crystal. As can be seen in Table \ref{table:XSWresults}, the Bi and Se coherent positions differ only slightly from the values obtained for the \BiSe{} crystal, while the coherent fractions are smaller for \SrBiSe. We conclude that the crystalinity of the doped crystal is worse compared to the undoped crystal, but certainly sufficient for a meaningful NIXSW experiment. For the following analysis we assume that the inelastic mean free path that was determined for the undoped crystal ($\lambda = 65$~\AA, see above) is unchanged by the doping. 

In Fig.~\ref{fig:Sr-Argand}, the calculated NIXSW results (Argand vectors) for Sr are shown, considering all possible Sr positions (heights) in the \BiSe{} unit cell, according to the symmetry of the structure. The full \BiSe{} unit cell exhibits inversion symmetry. Since in our NIXSW experiment we are only sensitive to vertical coordinates, all species effectively fulfill a mirror symmetry with two mirror planes, one located at the height of the central Se layer in each QL (Se3, 8, ...), the other in the center of the vdW gap. Hence, any arbitrarily selected (vertical) site for a dopant species is duplicated within one QL by the first mirror plane, and multiplied into all other QLs by the second. This is illustrated in Fig.~\ref{fig:Sr-Argand}(a) by the color coding from dark blue to dark red (see color-bar in the left), representing all equivalent heights between the center of the vdW gap (blue) and the Se3 layer in the middle of the QLs (red). The color scale is multiplied into every other half of a QL according to the mirror planes. Cyan spheres are drawn to illustrate one specific case, a Sr position close to the Se1 and Se5 planes. As discussed in the following, this is the principal dopant position we found in our NIXSW results. 

In Fig.\ \ref{fig:Sr-Argand}(b) we show the simulated NIXSW result \cite{Mercurio2014} for all possible dopant positions. The rainbow-colored spiral-type line indicates the simulation results (the corresponding Argand vectors end on the line) for varying the dopant position, whereby the same color code as in Fig.~\ref{fig:Sr-Argand}(a) was used. All multiplications of the specific sites (heights) by the mirror symmetry are considered and weighted with their corresponding attenuation factor. The shape of the line can be understood in the following gedankenexperiment: 

We start positioning the Sr dopants in the center of the vdW gap (dark blue color). The corresponding calculated Argand vector is $(\Fc, \Pc) = (1.0, 0.5)$, as indicated by the dark blue color of the line at this position in Fig.~\ref{fig:Sr-Argand}(b). When we now move the Sr atom through the crystal towards the Se3 plane, the expected Argand vector moves through the upper half of the diagram along the blue line section towards the right, at first reaching $(\Fc, \Pc) = (1.0, 0.0)$, close to the position of the Se1 layer. Between Se1 and Bi2 it moves back along the lower blue-cyan-green line section to the starting point, and so on along the green-yellow-orange-red path. One full turn of the line (e.g., from $(1.0, 0.0)$ through (1.0, 0.5) and back to $(1.0, 0.0)$), corresponds to moving the Sr atom from one Bragg-plane to the next, a distance that approximately equals the distance between two neighboring layers. The fact that the spiral-type line flattens from blue to red reflects the fact that the two equivalent Sr atoms in one QL move closer to each other, i.e., the difference in photoelectron yield from atoms on both sites becomes smaller due to a reduced inelastic mean free path effect. This reduces the resulting $\Fc$ after averaging over all equivalent sites. Note that all Argand vectors would end on the line connecting $(\Fc, \Pc) = (1.0, 0.0)$ and $(1.0, 0.5)$ (the spiral-type line collapses to straight horizontal lines), if the inelastic mean free path effect was neglected ($\lambda \rightarrow \infty$). 

It can clearly be seen that our experimental data (black data points and black radial line) exclusively agrees with the position color coded in cyan, corresponding to a dopant position close to those Se layers lying next to the vdW gap (Se1, 5, 6, 11, ...). The best result is obtained for $(\Fc, \Pc) = (0.41, 0.86)$,  
for a discussion of reliability see Sec.\ \ref{sec:Appendix} (Appendix).
This corresponds to a distance of only  $0.08$~\AA\ between the Sr atoms and the Se1/Se5 layers (Sr closer to the center of the QL). The corresponding distance between the Sr atoms and the central Se layer (Se3) of the QL is $3.42$~\AA, see Fig.\ \ref{fig:Sr-Argand}. 

It is worth mentioning that our simulation is model-free. The only parameter used is the inelastic mean free path of the escaping photoelectrons, which we have fixed to the value obtained from the measurements on the \BiSe{} cystal ($\lambda = 65$~\AA). Changing $\lambda$ would stretch the spiral-type line vertically. However, in order to move the next closest calculated positions, the yellow-orange part of the eliptical curve, close to our experimental data points, we had to decrease the inelastic mean free path by a factor of $\approx 3$, which appears to be very unrealistic. This result would even hold if a propper correction of the Sr 3d yield for non-dipolar effects was possible \cite{NDP}, since the change in $\Fc$ caused by such a correction is expected to be less than $20$\% \cite{vanStraaten2018}.

Finally, we would like to address the aspect of surface relaxation in a similar way as explained above for the bulk species Se and Bi. We simulated how the expected NIXSW result will change if we allow a relaxation of the uppermost Sr layer by $\Delta_{Sr1} = \pm 0.3$~\AA, while keeping the position of all other Sr atoms fixed. The result is plotted as a gray line in Fig.~\ref{fig:Sr-Argand}(b) running through the point $(\Fc, \Pc) = (0.41, 0.86)$, that is the center of the region of good agreement between simulation and experiment (thick cyan line). Close to that point (in particular towards the left), the gray line follows closely the cyan part of the spiral-type line, before it starts deviating to smaller coherent fractions. This indicates that -- in a certain range and to some extent -- a relaxation of the uppermost Sr atoms can compensate for a slightly different position of all other Sr atoms in the QL, i.e., these parameters are highly correlated in a certain region. This correlation induces a certain uncertainty in our analysis, which, however, can be quantified: We define the two relevant parameters as $d_{Sr}$, the distance of all Sr atoms to the center of the QL, i.e., to the Se3 layer, and $\Delta_{Sr1}$ as the (additional) relaxation of the uppermost Sr atoms (those located close to the Se1 layer), with positive values corresponding to an outward relaxation, i.e., to an increase of the distance to the Se3 layer. 

As mentioned above, we have obtained the best agreement of experiment and simulation for $d_{Sr} = 3.42$~\AA\ (with $\Delta_{Sr1}=0$), and simulated the relaxation in the range of $\Delta_{Sr1} = \pm 0.3$~\AA\ (gray line 
in Fig.\ \ref{fig:Sr-Argand}(b)). The range, in which these two parameters are highly correlated, is mainly located  in the left half of the gray line. On the right side, the line almost immediately deviates significantly from the cyan colored line section. Hence, the interval of correlation is asymmetric, and can be estimated to  $\Delta_{Sr1} = [-0.2, 0.1]$~\AA. 

The corresponding change of the overall Sr position, which can be compensated by such a relaxation $\Delta_{Sr1}$, is significantly smaller. This is the case simply because a change of $d_{Sr}$ changes the position of many Sr atoms in the crystal, with a much greater effect on the coherent position than the displacement of only the Sr1 atoms. The corresponding interval for a change of $d_{Sr}$ is therefore only $[+0.07, -0.03]$~\AA\ (note the opposite sign of the numbers compared to the $\Delta_{Sr1}$ interval). In other words, a change of the overall position of the Sr atoms by $+0.07$~\AA, that is from $d_{Sr} = 3.42$~\AA\ to $d_{Sr} = 3.49$~\AA, can be compensated by a relaxation of the uppermost Sr atom by $\Delta_{Sr1} = -0.2$~\AA. In the opposite direction, the corresponding shifts are smaller. A change of $d_{Sr1}$ by $-0.03$~\AA\ is compensated by $\Delta_{Sr1} = +0.1$~\AA. These numbers represent the \textit{maximum} shifts of both parameters that can compensate each other, and hence quantify the uncertainty of our result caused by the correlation of these two parameters. The (additional) statistical uncertainty in our measurements is estimated to $\pm 0.03$~\AA.

\section{Conclusions}

We have performed a combined STM and NIXSW study on undoped and Sr-doped \BiSe\ samples. The STM measurements resolved different native defects, namely Bi-antisites (Bi$_\text{Se}$), Se-interstitials (Se$_\text{int}$), and Se vacancies (V$_\text{Se}$), in the top quintuple layer, but did not show any atomic scale defect that was unique to the \SrBiSe{} crystals. 
A quantitative comparison of the defect densities found in the STM data showed an increase of Bi-antisite defects in the Se1 and Se5 layers in \SrBiSe{} compared to \BiSe\ by about one order of magnitude.
This increase may be related indirectly to Sr-dopants, i.e., Sr atoms replacing Bi atoms and Bi atoms in turn replacing Se atoms, thereby creating additional Bi antisites. 
While Sr substitution at the Bi sites is energetically favorable \cite{Li2018}, the observed Bi antisites density can only account for about 4\% of the expected nominal Sr dopant concentration, and hence the majority of Sr dopants must occupy other sites which did not show up in the surface areas probed by our STM measurements.
This apparent disagreement may be explained by the known inhomogeneity in these kind of samples when it comes to dopant concentration~\cite{Kriener2011,Sobczak2018}.

Since Sr dopants could not be identified based on a characteristic appearance in STM images, we also performed NIXSW measurements on both types of samples. NIXSW is an ideal tool to determine the dopant sites in the (near-surface) bulk crystal unambiguously. The method is representative for a larger sample volume, since it integrates over several thousand square-micrometers at the surface and has a probing depth of several QLs, sufficient for our purpose. Due to the depth-dependent attenuation of the photoelectrons (inelastic mean free path), it has the resolution to distinguish between otherwise symmetry-equivalent dopant sites in the bulk. The latter causes an apparent break of the structure's symmetry, and constitutes the backbone of our analysis.

The main result of our NIXSW study on the \SrBiSe{} sample is the unambiguous identification of the dopant vertical position: The majority of the Sr dopants is located close to the Se layers Se1 and Se5, vertically displaced towards the center of the QL by only $(0.08 ^{-0.10}_{+0.06})$~\AA. This corresponds to a distance of $d_{Sr} = (3.42^{+0.10}_{-0.06})$~\AA\ to the Se3 layer in the center of the QL. In particular, this means that any significant amount of Sr atoms within the vdW gap can be excluded unambiguously by our analysis. If there are dopants in the vdW gap, there occupation is minor and not detectable in NIXSW. The same is valid for the positions closer to the center of the QL, i.e., close to or between the Bi2, Se3 and Bi4 layers. Note that our NIXSW is only sensitive to vertical positions of all species, and hence the lateral position of the dopants cannot be resolved. However, steric and chemical bonding conditions suggest that the Sr dopants occupy Se lattice sites in the Se1 and Se5 layers, only slightly relaxed towards the center of the QL.

When comparing these experimental result for the Sr dopant site with previous reports, we find some agreement with the work by Li \textit{et al.} \cite{Li2018}: These authors also found Sr at similar positions between the Se1 and Bi2 layer, but \textit{not} within the vdW gap. However, the same data also showed some Sr atoms in between all other layers within the QLs. 
Interestingly, density functional theory calculations by the same group showed that Sr in between Se1 and Bi2 migrates into the vdW gap when the system is allowed to structurally relax. 
This apparent contradiction is however remedied by the need of quenching the crystals at the end of the crystal growth procedure. Apparently, Sr atoms are, in this way, trapped in metastable adsorption sites, and therefore, these energetically unfavorable dopant sites may indeed be predominantly responsible for the high superconducting volume fraction observed in quenched crystals \cite{Shruti2015}.

\acknowledgments{We thank Diamond Light Source for access to beamline I09 (via proposal SI23317), and the I09 beam-line staff (P. K. Thakur, D. Duncan, and D. McCue) for their support during the experiment. F.C.B., C.K. and F.S.T. acknowledge funding by the Deutsche Forschungsgemeinschaft (DFG, German Research Foundation) through SFB 1083 "Structure and Dynamics of Internal Interfaces", subproject A12. The work at Cologne has received funding from the European Research Council (ERC) under the European Union's Horizon 2020 research and innovation programme (grant agreement No 741121) and was also funded by the DFG under CRC 1238 - 277146847 (Subprojects A04 and B06) as well as under Germany's Excellence Strategy - Cluster of Excellence Matter and Light for Quantum Computing (ML4Q) EXC 2004/1 - 390534769.
}

\section*{Author contributions}
Y.-R. L. and M. B. contributed equally to this work.
Y. A. and F. S. T. conceived the research.
M. B. grew the crystals and studied them using SQUID.
M. B. and J. B. performed and analyzed the STM experiments.
Y.-R. L., M. B., S. S., T.-L. L., J. B., F. C. B. and C. K. performed the NIXSW experiments, and 
Y.-R. L., F. C. B. and C. K. analyzed the data.
Y.-R. L., M. B., F. C. B. and C. K. made the figures, and 
J. B., F. C. B. and C. K. wrote the paper (with some input from all other authors).

\section{Appendix}
\label{sec:Appendix}

\begin{figure}[t]
	\centering
	\includegraphics[width=0.98\columnwidth]{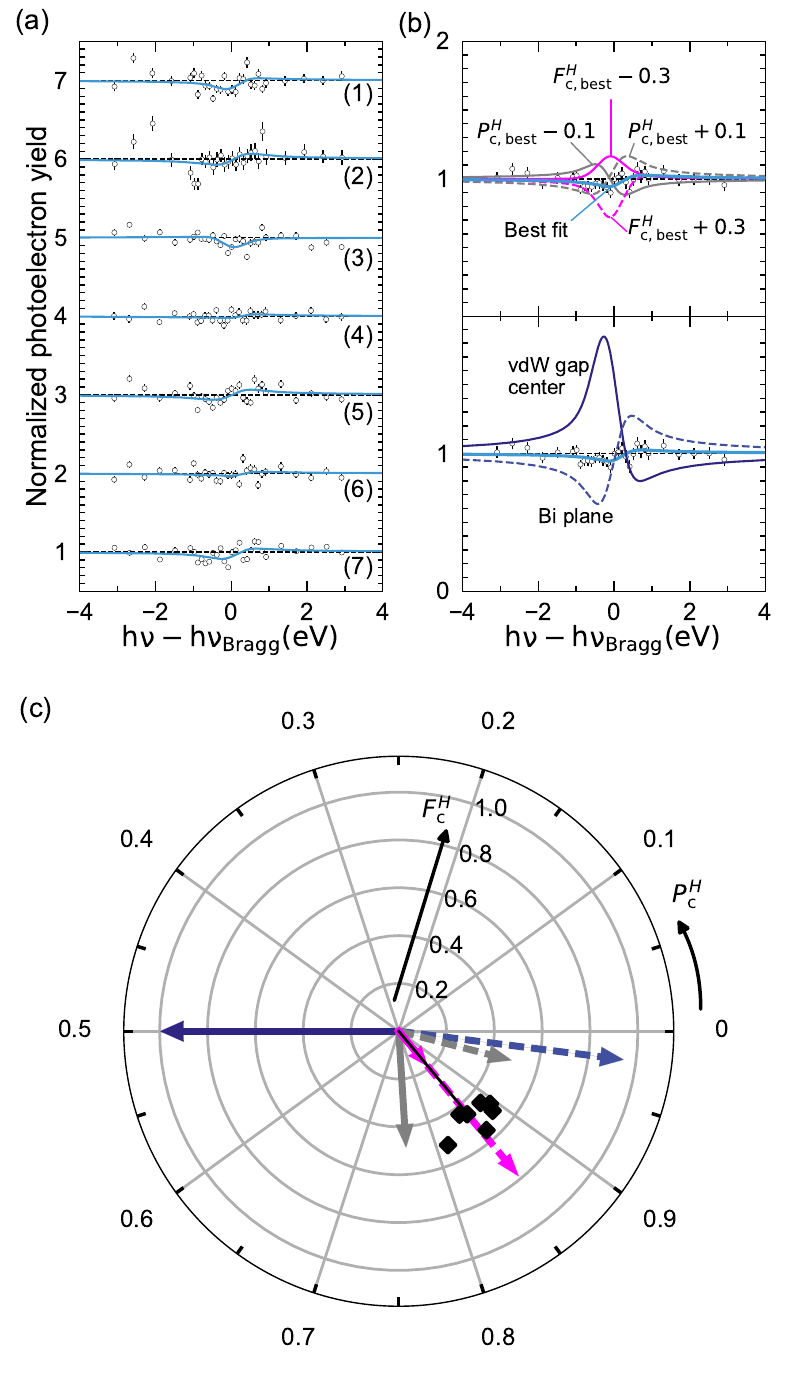}
	\caption{     \label{fig:Sr-XSW-appendix}
		Full Sr 3d NIXSW data set. (a) Data of all seven individual scans (open black circles) and best fit curves (cyan line). Scans (1) and (2) were measured with slightly different settings than (3-7). Scans (1-6) are vertically displaced for better visibility. (b) Averaged data (scans 3-7) and best fit, as shown in Fig.\ \ref{fig:XPS-yield}. Additionally, in the upper part, calculated fit curves for higher and lower coherent positions (gray dashed and solid lines, respectively), as well as higher and lower coherent fractions (magenta dashed and solid lines, respectively) are shown. In the lower part, calculated yield curves for Sr positioned in the center of the vdW gap and in the Bi2/4 plane are shown (solid and dashed blue line, respectively). (c) Argand diagram with $(\Fc, \Pc)$ vectors corresponding to the yield curves shown in (b), using the same color code. 
	}
\end{figure}

As can be seen in Fig.\ \ref{fig:XPS-yield}, the statistics of the Sr data set is comparably poor compared to Se and Bi, owing to the small doping concentration of Sr in the doped \BiSe\ crystal, and the yield curve is comparably flat and unstructured. Nevertheless, here we demonstrate that the data analysis is reliable and yields unambiguous results.

In Fig.\ \ref{fig:Sr-XSW-appendix}(a) we show all seven individual NIXSW scans which we have performed on the Sr 3d core level. The data collection time was up to $70$~min per scan. The normalized electron yield is uniformly flat in all scans. Considering the theoretical description of the yield by Eq.\ \ref{eq:yield}, this means that the second and third term, i.e., the reflectivity $R(h\nu)$ and the interference term $\sqrt{R} \cos(...)$, are cancelling out. This is obviously only the case for a coherent fraction significantly higher than zero, since otherwise (with $F_c^H=0$) the interference term would vanish and the yield curve would simply follow $1+R(h\nu)$. A coherent fraction above zero, in turn, also implies a meaningful coherent positon. In other words, although a flat yield curve might naively be mistaken for the opposite, it in fact represents a well defined situation and indicates that well interpretable results can be obtained for coherent fraction and position. The good agreement of seven independent measurements strengthens this fact. 

We deepen this claim in the following by showing a number of calculated yield curves for scenarios that differ from our best-fit result. First, starting from our best fit (thick cyan line in Fig.\ \ref{fig:Sr-XSW-appendix}(b)), we change the coherent position by $\pm 0.1$ (gray solid and dashed lines in the upper part of Fig.\ \ref{fig:Sr-XSW-appendix}(b), calculated for $\Pc = 0.76$ and $0.96$, respectively). These differences in the coherent position correspond to a change of the vertical Sr position of approx.\ $\pm 0.3$\AA, that is only $10$\% of the interlayer spacing. In Fig.\ \ref{fig:Sr-XSW-appendix}(c) the corresponding Argand vectors are displayed using the same line colors and styles. Second, we change the coherent fraction by $\pm 0.3$ (magenta solid and dashed line in Fig.\ \ref{fig:Sr-XSW-appendix}(b)). All of these four calculated curves fit our data significantly worse than our best fit, and demonstrate the high sensitivity of our data set for the vertical position of the Sr species. 

Note that the solid magenta yield curve is calculated for $(\Fc, \Pc) = (0.19, 0.86)$, corresponding to an Argand vector ending on the yellow-orange part of spiral-type line in Fig.\ \ref{fig:Sr-Argand}.  This is the part of the simulation coming closest to our experimental data, beside the cyan part. The clear misfit of this solid magenta line with our data shows that such a scenario represented by the yellow-orange color code, that is a Sr position in between the Bi2/4 and Se3 planes, can be clearly excluded.

The same is valid for a Sr position close to the center of the vdW gap, $(\Fc, \Pc) = (0.5, 1.0)$. The corresponding yield curve and Argand vector are shown with solid dark-blue lines in the bottom part of Fig.\ \ref{fig:Sr-XSW-appendix}(b) and in (c). And finally we simulated the yield curve for Sr atoms substituting Bi in their Bi2/4 planes. The yield curve representing this scenario with $(\Fc, \Pc) = (0.95, 0.98)$ is shown as a dashed blue line in the bottom part of Fig.\ \ref{fig:Sr-XSW-appendix}(b), and it is obvious that it can also not fit our experimental data. 

In conclusion, this detailed analysis of our Sr NIXSW data and the presented model calculations unambiguously confirm the conclusion drawn in the main text: The Sr 3d NIXSW data can only be fitted by a model positioning the Sr dopants close to the Se1/5 planes.

\end{document}